\documentclass[twocolumn,nofootinbib,prd]{revtex4}
\usepackage{nicefrac}
\usepackage{latexsym}
\usepackage{amsmath, mathrsfs, bm}
\usepackage{eqnarray}
\usepackage{url}
\usepackage{cases}
\usepackage{epsfig}
\usepackage{color}
\usepackage{graphicx,subfigure}
\usepackage{grffile}
\usepackage{float}
\usepackage{natbib}
\usepackage{soul}
\usepackage{lmodern}

\usepackage{times}

\usepackage{etoolbox}
\preto\align{\par\nobreak\small\noindent}
\expandafter\preto\csname align*\endcsname{\par\nobreak\small\noindent}
\preto\multline{\par\nobreak\small\noindent}
\expandafter\preto\csname align*\endcsname{\par\nobreak\small\noindent}
\preto\flalign{\par\nobreak\small\noindent}
\expandafter\preto\csname align*\endcsname{\par\nobreak\small\noindent}
\preto\eqnarray{\par\nobreak\small\noindent}
\expandafter\preto\csname align*\endcsname{\par\nobreak\small\noindent}

\newcommand{\citepy}[1]{\citeauthor{#1}~\cite{#1}}
\newcommand{\reffg}[1]{Fig.~\ref{#1}}
\newcommand{\reftb}[1]{Table~\ref{#1}}
\newcommand{\refeq}[1]{Eq.~(\ref{#1})}
\newcommand{\refsc}[1]{Sec.~\ref{#1}}

\newcommand*{\msk}{\\[0.25cm]} 
\newcommand*{\nmsk}{\notag\msk} 

\newcolumntype{L}[1]{>{\raggedright\arraybackslash}p{#1}}
\newcolumntype{C}[1]{>{\centering\arraybackslash}p{#1}}  
\newcolumntype{R}[1]{>{\raggedleft\arraybackslash}p{#1}}

\def\planck{{\it Planck~}}

\begin{document}

\title{
    Measurement of the pairwise kinematic Sunyaev-Zeldovich effect
    with \textit{Planck} and BOSS data}

\author{\textsc{Yi-Chao Li}$^{1,2}$}
\author{\textsc{Yin-Zhe Ma}$^{1,2,3}$}
\email{Ma@ukzn.ac.za}
\author{\textsc{Mathieu Remazeilles}$^{3}$}
\author{\textsc{Kavilan Moodley}$^{4,2}$}
\affiliation{
    $^1$ School of Chemistry and Physics, University of KwaZulu-Natal, 
    Westville Campus, Private Bag X54001, Durban 4000, South Africa \\
    $^{2}$NAOC-UKZN Computational Astrophysics Centre (NUCAC), 
    University of KwaZulu-Natal, Durban, 4000, South Africa \\
    $^3$ Jodrell Bank Centre for Astrophysics, Alan Turing Building, 
    School of Physics \& Astronomy, The University of Manchester,
    Oxford Road, Manchester, M13 9PL, U.K \\
    $^{4}$Astrophysics and Cosmology Research Unit, School of Mathematics.
Statistics and Computer Science, University of KwaZulu-Natal, Durban, 4041, South Africa
}


\begin{abstract}
    We present a new measurement of the kinetic Sunyaev-Zeldovich effect (kSZ) using
    {\it Planck} cosmic microwave background (CMB) and Baryon Oscillation
    Spectroscopic Survey (BOSS) data. Using the `LowZ North/South' galaxy catalogue from
    BOSS DR12, and the group catalogue from BOSS DR13, we evaluate the mean pairwise kSZ
    temperature associated with BOSS galaxies. We construct a `Central Galaxies Catalogue'
    (CGC) which consists of isolated galaxies from the original BOSS data set, 
    and apply the aperture photometry (AP) filter to suppress the primary CMB 
    contribution. By constructing a halo model to fit the pairwise kSZ function,
    we constrain the mean optical depth to be 
    $\bar{\tau}=(0.53\pm0.32)\times10^{-4}(1.65\,\sigma)$ for `LowZ North CGC',
    $\bar{\tau}=(0.30\pm0.57)\times10^{-4}(0.53\,\sigma)$ for `LowZ South CGC', 
    and 
    $\bar{\tau}=(0.43\pm0.28)\times10^{-4}(1.53\,\sigma)$ for `DR13 Group'.
    In addition, we vary the radius of the AP filter and find that the AP size
    of $7\,{\rm arcmin}$ gives the maximum detection for $\bar{\tau}$.
    We also investigate the dependence of the signal with halo mass and find
    $\bar{\tau}=(0.32\pm0.36)\times10^{-4}(0.8\,\sigma)$ and
    $\bar{\tau}=(0.67\pm0.46)\times10^{-4}(1.4\,\sigma)$ 
    for `DR13 Group' with halo mass restricted to, respectively, less and greater than
    its median halo mass, $10^{12}\, h^{-1}{\rm M}_\odot$.
    For the `LowZ North CGC'  sample restricted to
    $M_{\rm h} \gtrsim 10^{14}\, h^{-1}{\rm M}_\odot$ there is no detection 
    of the kSZ signal because these high mass halos are associated with the high-redshift
    galaxies of the LowZ North catalogue, which have limited contribution to the 
    pairwise kSZ signals. 
\end{abstract}

\maketitle

\section{introduction}

The kinematic Sunyaev-Zeldovich effect (hereafter, kSZ effect), 
first proposed in \cite{1972CoASP...4..173S,1980MNRAS.190..413S},
describes the anisotropy of the CMB due to its scattering off moving electrons
in the Universe
\begin{align}
    \frac{\Delta T}{T}=-\frac{\sigma_{\rm T}}{c}
    \int {\rm d}l\, n_{\rm e}\,(\mathbf{v}\cdot \hat{\mathbf{n}}),
\end{align}
where $\sigma_{\rm T}$ is the Thomson cross section, $n_{\rm e}$ is the electron 
density and
$\mathbf{v}$ is the peculiar velocity of electrons relative to the CMB. 
In the limit of nonrelativistic elastic scattering (Thomson scattering), 
the kSZ effect is equally efficient for all frequencies and causes a
frequency-independent distortion of CMB spectrum. 

In recent years, there have been a series of works to detect and measure the kSZ signal.
\citepy{2012PhRvL.109d1101H} first reported the detection of the kSZ signal by applying the
pairwise kSZ estimator~\cite{1999ApJ...515L...1F} to CMB data from the Atacama
Cosmology Telescope (ACT~\cite{2011ApJS..194...41S}) using a galaxy catalog from the Sloan
Digital Sky Survey III DR9~\cite{2012ApJS..203...21A}. Recently, this measurement of
the pairwise kSZ effect was achieved with higher precision using ACT CMB data combined with the BOSS DR11
catalogue~\cite{2017JCAP...03..008D}. With the \planck map, \citepy{2016A&A...586A.140P}
reported a kSZ detection using the pairwise kSZ estimator on the Central Galaxy Catalogue (CGC) extracted from SDSS DR7~\cite{2005AJ....129.2562B}.
Reference~\cite{2016A&A...586A.140P} also constructed the correlation function between the 
peculiar velocity field and kSZ temperature anisotropies and achieved the first detection at
$3.0\,\sigma$ confidence level (C.L.).  Apart from the use of spectroscopic surveys,
\citepy{2016MNRAS.461.3172S} detected the pairwise kSZ effect with photometric survey data
from the Dark Energy Survey (DES~\cite{2016MNRAS.460.1270D}) and CMB data from the
South Pole Telescope (SPT\cite{2015ApJ...799..177G}). 

In addition to the pairwise temperature difference estimator of the kSZ effect,
there have been several other statistical methods developed to measure the kSZ effect.
\citepy{2016JCAP...07..001S} proposed the density-weighted pairwise kSZ estimator
in Fourier space, and achieved a $2.54 \, \sigma$ detection 
with the \planck CMB map and BOSS DR12 catalogue\cite{2017arXiv170507449S}.
\citepy{Hill16} and \citet{Ferraro16} cross-correlated the squared temperature
map from \planck and {\it Wilkinson Microwave Anisotropy Probe} 
({\it WMAP}~\cite{2013ApJS..208...20B}) data, 
appropriately filtered to isolate the small-scale kSZ signal,
with the projected galaxy positions from the 
{\it Wide-field Infrared Survey Explorer} 
({\it WISE}~\cite{2010AJ....140.1868W}) and achieved a 
$\sim 4.0\,\sigma$ C.L. detection of the kSZ effect. More recently, 
\citepy{Planck17-dispersion} detected the velocity dispersion effect of the kSZ
signal from the \planck {\tt 2D-ILC} CMB map at $3.2\,\sigma$ C.L.

As the measurement of the kSZ effect becomes more accurate, there is the potential to
use it to trace the peculiar velocities of galaxies and clusters. The pairwise kSZ 
signal is closely related to the pairwise velocity of the galaxies or clusters, 
which encodes information about large-scale structure. Previous studies showed 
that the kSZ measurement can
be used to constrain the dark energy equation of state
\cite{2005astro.ph.11060D,2006ApJ...643..598H,
2007ApJ...659L..83B,2008PhRvD..77h3004B,2014PhLB..735..402M,2015ApJ...808...47M} and
modified gravity models~\cite{2009PhRvD..80f2003K}. Recently, \citepy{2015PhRvL.115s1301H}
showed that the baryon fraction is consistent with unity for the kSZ--peculiar velocity
field cross-correlation signal. 
In addition, \citepy{2015PhRvD..92f3501M} pointed out that the
the kSZ effect can be used to constrain neutrino mass with 
precision measurements and large enough samples.

In this work, we focus on measuring the pairwise kSZ signals using the foreground-cleaned 
\planck {\tt 2D-ILC} CMB map, which, by construction, is free from thermal SZ (tSZ) 
residual contamination~\cite{2011MNRAS.410.2481R}, and the galaxy and group catalogues 
from BOSS DR12 and DR13. In Sec.~\ref{sec:data} we present the details of the CMB map and 
the galaxy and group catalogues that we use. The estimator and theoretical model are introduced 
in Sec.~\ref{sec:estimator} and Sec.~\ref{sec:model} respectively. We summarize and discuss our results 
in Sec.~\ref{sec:discuss}.  In our analysis, we adopt a spatially flat $\Lambda$CDM cosmology
 model with the cosmological parameters fixed to the best-fit values of~\citepy{2016A&A...594A..13P}: 
$\Omega_{\rm m}=0.309$, $\Omega_{\rm \Lambda}=0.691$, $n_{\rm s}=0.9608$, $\sigma_8=0.815$ and $h=0.68$.

\section{DATA}\label{sec:data}

\subsection{The {\textbf{\textit{Planck}}} {\tt 2D-ILC} CMB map}

In this work we use the {\tt 2D-ILC} CMB map, which we have obtained by applying the `Constrained ILC' component separation method~\cite{2011MNRAS.410.2481R}
to the public \planck 2015 data\footnote{\url{http://pla.esac.esa.int/pla}}. 
The `Constrained ILC' method is specifically designed to nullify the residuals of thermal Sunyaev-Zeldovich (tSZ) emission in the CMB map, while minimizing 
the contamination from other foregrounds and noise. Along similar lines to those of the standard NILC method~\cite{Delabrouille2009,2012MNRAS.419.1163B}, the constrained ILC approach performs a minimum-variance weighted linear combination of the nine \planck frequency maps, with the weights giving unit response to the CMB spectral energy distribution. However, the `Constrained ILC' method 
offers an additional constraint for the vector of weights to be orthogonal to the spectral energy distribution of the tSZ effect, thus providing a null response to the tSZ signal during the filtering. As a result, the so-called {\tt 2D-ILC} CMB map, where 2D stands for the two-dimensional constraint on CMB and tSZ, is free from tSZ residuals in the direction of the galaxy clusters, unlike other CMB maps released by \emph{Planck}. Therefore, the {\tt 2D-ILC} CMB map allows us to probe the kSZ effect in the direction of galaxy clusters without suffering from tSZ bias and variance. 
 The full-width-half-maximum (FWHM) of the {\tt 2D-ILC} map is
$\theta_{\rm FWHM}=5\,$arcmin. For more details on the `Constrained ILC' method and on the {\tt 2D-ILC} CMB map, we refer to~\cite{2011MNRAS.410.2481R} and Sec.~2.1 of~\citepy{Planck17-dispersion}.

\subsection{Catalogue of galaxies}

\begin{figure} 
    \centering
    \small
    \includegraphics[width=0.45\textwidth]{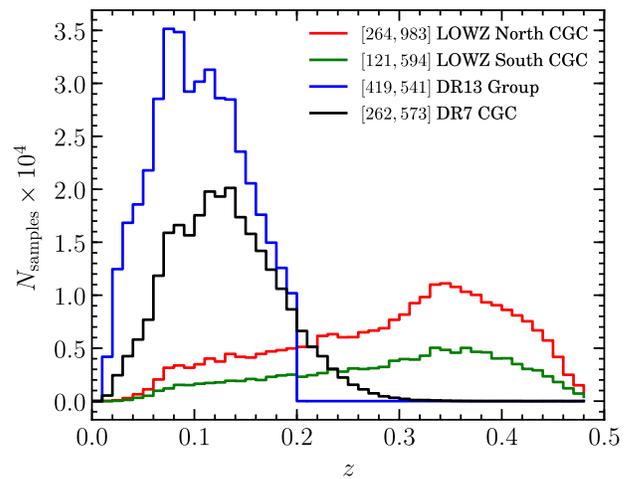}
    \caption{
        The redshift histogram
        for each catalogue. 
        The total number of galaxies in 
        each catalogue is shown in the legend. The DR7 CGC catalogue
        that was used in \citepy{2016A&A...586A.140P} is also shown here
        as the black curve. 
    }\label{fig:zhist}
\end{figure}

\begin{table}
    \centering
    {\scriptsize
    \caption{Summary of the catalogue information}\label{tab:cata-info}
    \begin{tabular}{c|c|c|c} \hline\hline
        Catalogue
        & LowZ North CGC
        & LowZ South CGC
        & Group DR13 \\ \hline

        $N_{\rm sample}$
        & 264,983
        & 121,594
        & 419,541
        \\

        $S_{\rm overlap}[{\rm deg}^{2}]$
        & $\sim 5600$
        & $\sim 2500$
        & $\sim 6900$
        \\

        $z$
        & $[0.01, 0.50]$
        & $[0.01, 0.50]$
        & $[0.01, 0.20]$
        \\

        $\log_{10} \left(M_{\rm h}/h^{-1}{\rm M}_{\odot}\right)$
        & $[10, 19]$
        & $[10, 19]$
        & $[11, 15]$
        \\ \hline\hline
    \end{tabular}
    }
\end{table}

\paragraph{Central galaxy catalogue}
The galaxy samples in the Central Galaxy Catalogue (CGC) are selected to 
trace the centres of dark matter halos. We follow the method used in
Ref.~\cite{2016A&A...586A.140P} to find isolated galaxies with no other 
galaxies within $1.0\,{\rm Mpc}$ in the transverse direction
and with a redshift difference smaller than $1000\,{\rm km}\,{\rm s}^{-1}$.
Such an isolation criterion is applied to the large-scale structure 
galaxy catalogue of the twelfth data release of
the Baryon Oscillation Spectroscopic Survey (BOSS DR12).
The BOSS sample was designed to measure the BAO signature in the 
two-point galaxy clustering statistical analysis, and separated into
LowZ and CMASS (for `Constant (stellar) Mass') catalogues
\cite{2016MNRAS.455.1553R}. Both LowZ and CMASS include
two separated survey areas, located in the Northern and Southern Galactic caps,
respectively. We use `LowZ North/South' as the
labels for the two different LowZ subcatalogues. 
After applying the isolation criterion, around $20\%$ of the galaxies are flagged out
for `LowZ North/South' catalogue.
The redshift histograms for the `LowZ North/South CGC' are
shown in \reffg{fig:zhist}, and the  total number of galaxies in each catalogue
are listed in
the legend. Figure~\ref{fig:zhist} also shows the redshift distribution of
the CGC selected from SDSS DR7 with label `DR7 CGC', 
which is used in the kSZ analysis of Ref.\cite{2016A&A...586A.140P}.
After applying the CGC criterion, the sample amount of the `LowZ North CGC'
is comparable to that of `DR7 CGC',
but the `LowZ North CGC' samples, as well as the `LowZ South CGC' samples,
have broad redshift distributions with a median value of $z_{\rm median} = 0.315$.
The LowZ sample is designed to extend the SDSS-I/II Luminous Red Galaxy (LRG)
sample to higher redshift and fainter luminosities. But it also includes
a bright magnitude cut, which excludes a large number of the low-redshift galaxies.
We will discuss how the magnitude cut affects the kSZ signals in 
\refsc{sec:discuss}.

\paragraph{Group catalogue}
We also analyse the recently updated SDSS group catalogue 
\cite{2017MNRAS.470.2982L}, which is constructed with the galaxy samples
from the SDSS DR13 Northern Galactic Cap galaxy catalogue. The redshift
histogram of the SDSS group catalogue is shown as a blue line in~\reffg{fig:zhist}. The galaxies with
$z>0.2$ are cut out in this catalogue~\cite{2016A&A...586A.140P}. The group catalogue has a similar redshift 
distribution as the `DR7 CGC', which is the central galaxy catalogue
used in the previous analysis of \cite{2016A&A...586A.140P}.

The group finder used for the `DR13 Group' catalogue is based on that 
of Ref.~\cite{2005MNRAS.356.1293Y}, with improved halo mass assignment.
The improved group finder gives more accurate halo-mass estimates, 
and extends the group samples to even lower mass. 

The number of samples, sky coverage, halo mass and redshift range of LowZ North and South catalogue, and the DR13 group catalogue are shown in Table~\ref{tab:cata-info}.

\section{The Estimator}\label{sec:estimator}

The CMB brightness temperature fluctuation induced by the kSZ effect
of a galaxy cluster is given by
\begin{align}
    \delta T_{{\rm kSZ},i} = -\frac{T_0\bar{\tau}}{c} \mathbf{v}_i \cdot \hat{\mathbf{n}}_i,
\end{align}
where $\bar{\tau}$ is the mean optical depth of the sample, $T_0=2.725 {\rm K}$ is the mean
CMB temperature, $c$ is the speed of light, 
$\hat{\mathbf{n}}_i$ is the line-of-sight direction, and ${\mathbf{v}}_i$ is the peculiar velocity of the galaxy cluster relative to the CMB.
Here we make the assumption that the gas in the cluster traces the mass,
and that all clusters have the same gas-mass fraction. We also assume that the peculiar
velocity of free electrons is the
same as the velocity of dark matter, so that it traces the underlying dark matter distribution.

The kSZ effect of a single galaxy cluster
is known to be an order of magnitude lower than the tSZ effect and emission from dusty star-forming galaxies, thus the direct measurement
of the single-cluster kSZ effect is challenging. However,
the kSZ effect due to the relative movement between the cluster pairs has been
shown to be detectable
\cite{2012PhRvL.109d1101H,2017JCAP...03..008D,2016A&A...586A.140P}.
The pairwise momentum estimator is defined as~\cite{1999ApJ...515L...1F,2012PhRvL.109d1101H}
\begin{align}
    \hat{p}_{\rm kSZ}(r) = - \frac{\sum_{i<j}
    \left(\delta T_{{\rm kSZ},i} - \delta T_{{\rm kSZ},j}
    \right)c_{ij}}{\sum_{i<j}c^2_{ij}},
\end{align}
in which $i$ and $j$ indicate a pair of clusters, and $c_{ij}$ is the weight factor which only depends on the geometry
of the pair
\begin{align}
    c_{ij} = \frac{\left(r_i-r_j\right)\left(1 + \cos \theta\right)}
    {2\sqrt{r_i^2 + r_j^2 - 2r_ir_j\cos\theta}},
\end{align}
with $r_i$ and $r_j$ being the comoving distance of the two galaxies or clusters, and
$\theta$ being the angular separation between the two galaxies or clusters, i.e.
$\cos\theta=\hat{\mathbf{r}}_{i}\cdot \hat{\mathbf{r}}_{j}$.
If the sample size is large enough any signal that is independent of the comoving separation will be averaged to zero, due to the differential 
estimation. However, the infrared emission of the galaxies
and the tSZ signal may have small redshift-dependent variation, 
due to the cosmic evolution of the cluster mass and temperature.
The estimator can not avoid the contamination of the redshift-dependent
signals. Therefore, Refs.~\cite{2012PhRvL.109d1101H,2017JCAP...03..008D,2016A&A...586A.140P}
define the following Gaussian weighted average temperature of the sources as a function of
redshift, so that by subtracting it the redshift-dependent variation will be removed
\begin{align}
    {\mathcal T}(z) = \frac{
        \sum_i T_i \exp\left[-(z-z_i)^2/\sigma_z^2\right]}
    {\sum_i \exp\left[-(z-z_i)^2/\sigma_z^2\right]},
\end{align}
where $T_i$ is the temperature associated with the $i$-th sample
and $z_i$ is the redshift center of the $i$-th sample. 
The choice of $\sigma_z$ has no significant effect on the measurements~\cite{2012PhRvL.109d1101H,2017JCAP...03..008D,2016A&A...586A.140P}. In this work, we use $\sigma_z=0.001$. Then,
\begin{align}
    \delta T_{{\rm kSZ},i} = T_i - {\mathcal T}(z_i).
\end{align}
In order to extract $T_i$, we apply the aperture photometry (AP)
filter to the CMB map. The main advantage of the AP filter
is that it is independent of any assumptions of the halo profile.
The size of the AP filter can affect the detection of the kSZ 
pairwise momentum. The details of the AP filter are discussed further in \refsc{sec:discuss}.

\section{Model Fitting}\label{sec:model}

\subsection{Mean pairwise velocity}
The mean pairwise velocity between pairs of dark matter halos 
separated with comoving distance $r$ can be expressed with
the two-point correlation function of dark matter using 
the pair conservation equation
\cite{2008PhRvD..77h3004B,1977ApJS...34..425D},
\begin{align}\label{eq:pv}
    v(r, a(z)) = -\frac{2}{3}H(a)af(a)\frac{r\bar{\xi}^{\rm halo}(r, a)}
    {1 + \xi^{\rm halo}(r, a)},
\end{align}
where $a$ is the scale factor of the universe, $\xi^{\rm halo}(r, a)$ 
is the two-point correlation function,
\begin{align}
    \xi^{\rm halo}(r, a) = \frac{1}{2\pi^2} \int_0^{\infty}
    {\rm d}k\, k^2 j_0(kr) P(k, a) b^{(2)}_{\rm halo}(k, a),
\end{align}
where $j_{0}$ is the zero order spherical Bessel function, and $P(k;a)$ is the linear matter
power spectrum at scale factor $a$. Here
$\bar{\xi}^{\rm halo}(r, a)$ is the two-point correlation function
averaged over a sphere of radius $r$,
\begin{eqnarray}
    \bar{\xi}^{\rm halo}(r, a) &=&  \frac{3}{2\pi^2r^3} \int_0^r {\rm d}r' \, r'^2 \nonumber \\
& \times &    \int_0^{\infty} {\rm d}k\, k^2 j_0(kr') P(k, a) b^{(1)}_{\rm halo}(k, a).
\end{eqnarray}
The halo bias factor, $b^{(1)}_{\rm halo}(k, a)$ and $b^{(2)}_{\rm halo}(k, a)$
are given by
\begin{align}
    b^{(q)}_{\rm halo}(k, z) =
    \frac{\int {\rm d}M\, M ({\rm d}n/{\rm d}M)
        b(M, z)^q W^2(kR(M))}
        {\int {\rm d}M\, M ({\rm d}n/{\rm d}M) W^2(kR(M))}.
\end{align}
We take the expression of $b(M, z)$ in Ref.~\cite{2002MNRAS.336..112M},
\begin{align}
    b(M, z) = 1 + \frac{\delta^2_{\rm c} - \sigma^2(M, z)}
    {\sigma^2(M, z)\delta_{\rm c}},
\end{align}
where $\delta_{\rm c}=1.686$, and $\sigma^2(M, z)$ is the {\it rms} fluctuation on a mass scale $M$ at redshift $z$
\begin{align}
    \sigma^2(M, z) = \frac{1}{2\pi^2} \int_0^{\infty}
    {\rm d}k\,k^2 P(k, z) W^2(kR(M)).
\end{align}

The halo mass function (HMF) takes the form
\begin{align}
    \frac{{\rm d}n}{{\rm d}M}(M, z) = f(\sigma)
    \frac{\bar{\rho}_{\rm m}(z)}{M}
    \frac{{\rm d}\ln\sigma^{-1}}{{\rm d}M},
\end{align}
in which $\bar{\rho}_{\rm m}(z) = \Omega_{\rm m}\rho_{\rm crit}(1 + z)^3$
is the mean matter density of the Universe at redshift $z$ and
$\rho_{\rm crit} = 2.775 h^2 \times 10^{11}\,{\rm M}_{\rm \odot}\,{\rm Mpc}^{-3}$ is the critical
density of the Universe today. The collapse fraction $f(\sigma)$ has different expressions for different models. Here we use the 
parameterized $f(\sigma)$ and the fitting values of the parameters given by~\citepy{2010ApJ...724..878T},
\begin{align}
    f(\sigma, z) &= \alpha\left[1 + \left(
    \beta(z)\frac{\delta_{\rm c}}{\sigma}\right)^{-2\phi(z)}\right] 
    \left(\frac{\delta_{\rm c}}{\sigma}\right)^{2\eta(z)} 
    \exp\left[-\frac{\gamma(z)\delta_{\rm c}^2}{2\sigma^2}\right], \nonumber \\
    \beta(z) & = \beta  (1 + z) ^ { 0.20} , \nonumber \\
    \phi(z)  & = \phi   (1 + z) ^ {-0.08} ,\nonumber \\
    \eta(z)  & = \eta   (1 + z) ^ { 0.27} ,\nonumber \\
    \gamma(z)& = \gamma (1 + z) ^ {-0.01},
\end{align}
where $\alpha=0.368$, $\beta=0.589$, $\gamma=0.864$, $\phi=-0.729$ and $\eta=-0.243$~\cite{2010ApJ...724..878T}.

The pairwise velocity modeled by \refeq{eq:pv} is based on linear 
perturbation theory  and several assumptions, 
such as pair conservation of galaxies (halos), 
isotropic peculiar velocities, and isotropic clustering of galaxies (halos),
which are not valid in redshift space.
Ref.~\cite{2017arXiv170507449S} pointed out that the Kaiser effect 
from the redshift space distortions (RSDs) increases the amplitude of the 
pairwise kSZ signal by $\sim20\%$. Moreover, at nonlinear scales ($r<20\,h^{-1}$Mpc), 
the sign of the pairwise velocity function changes from negative to positive 
at a scale around $10\,h^{-1}\,{\rm Mpc}$~\cite{2014JCAP...05..003O,2016JCAP...07..001S,
2017JCAP...03..008D}. Therefore in our analysis, we restrict our model fitting to scales
$>20\,h^{-1}\,{\rm Mpc}$.

\subsection{Average optical depth}

The mean pairwise momentum is related to the mean pairwise velocity 
by a scaling factor,
\begin{align}
    p_{\rm kSZ}(r, a) = \bar{\tau} \frac{T_{\rm CMB}}{c} v(r, a),
\end{align}
where $\bar{\tau}$, as a free parameter, can be interpreted as the
mean optical depth within the solid angle of the AP filter size, 
averaged over the cluster samples.
We perform a $\chi^2$ minimization to find the best-fitting values for 
$\bar{\tau}$,
\begin{align}\label{eq:chisq}
    \chi^2 = \sum_{i,j} 
    \left(p_{\rm kSZ}^{\rm est}(r_i) - \bar{\tau}p_{\rm kSZ}^{\rm th}(r_i)\right)
    C^{-1}_{ij}
    \left(p_{\rm kSZ}^{\rm est}(r_j) - \bar{\tau}p_{\rm kSZ}^{\rm th}(r_j)\right),
\end{align}
where $p_{\rm kSZ}^{\rm est}(r_i)$ is the estimated kSZ momentum in the
$i$th separation bin, $p_{\rm kSZ}^{\rm th}(r_i)=(T_{\rm CMB}/c)v(r_i)$ 
is the theoretical predictions of the kSZ momentum (Eq.~(\ref{eq:pv})), and 
$C_{ij}$ denotes the $ij$ component of the covariance matrix.
The estimation of covariance matrix is discussed in \refsc{sec:cov}.
The best-fit value of $\bar{\tau}$ and its associated error are:
\begin{align}
    \bar{\tau} &= \frac
    {\sum_{i,j} p_{\rm kSZ}^{\rm est}(r_i)C^{-1}_{ij}p_{\rm kSZ}^{\rm th}(r_j)}
    {\sum_{i,j} p_{\rm kSZ}^{\rm th}(r_i)C^{-1}_{ij}p_{\rm kSZ}^{\rm th}(r_j)}
    \nmsk
    \sigma_{\bar{\tau}}^2 &= \frac{1}
    {\sum_{i,j} p_{\rm kSZ}^{\rm th}(r_i)C^{-1}_{ij}p_{\rm kSZ}^{\rm th}(r_j)}.
\end{align}

\section{Results \& Discussion}\label{sec:discuss}

\subsection{Correlation matrix and covariance matrix}\label{sec:cov}

\begin{figure}
    \small
    \centering
    \vspace{-0.5cm}
    \includegraphics[width=0.50\textwidth]{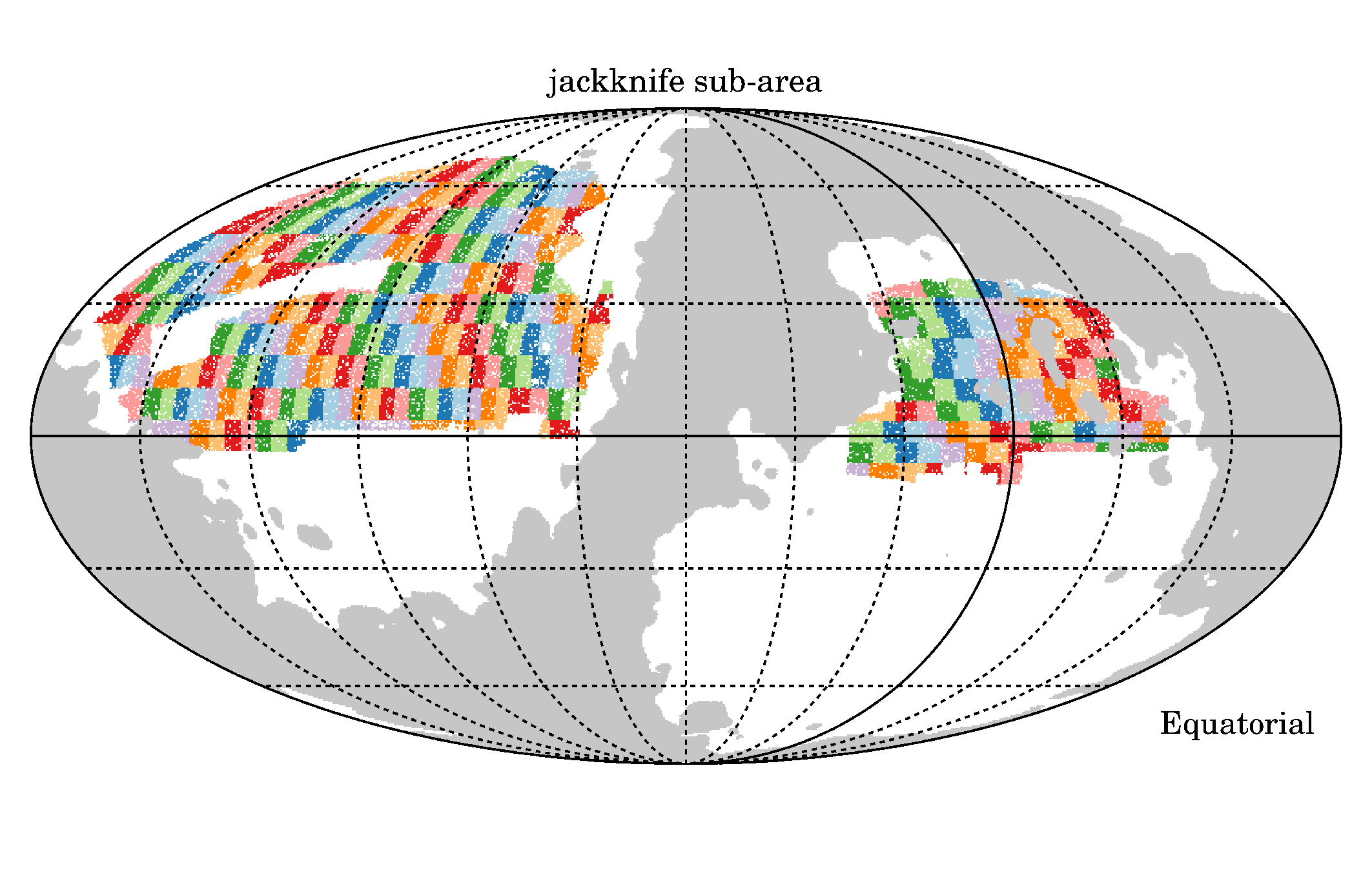}
    \vspace{-1cm}
    \caption{
        The jackknife subarea division for the `LowZ North' and `LowZ South'
        survey area shown in equatorial coordinates. The grey area shows the masked range of the \planck {\tt 2D-ILC} CMB map.
    }\label{fig:jksample}
\end{figure}

We first implement CMB mock samples including the thermal noise
to estimate the covariance matrix. The mock samples have the 
advantage that they are independent of each other and properly include
the cosmic variance contributed by the CMB. We generate $200$ 
mock maps with pure CMB fluctuations and thermal noise and run the estimator introduced above on each of the 
mock maps, using the source coordinates from the real catalogue.
Because there is no kSZ signal correlated with the sources, 
this method provides a {\it null} test for our estimator, and the estimated
covariance matrix has no contribution from the measurement variance of the 
signal.

Another method for estimating the covariance matrix is to use jackknife samples.
The jackknife sampling method \cite{1982jbor.book.....E} is widely used in the 
covariance matrix estimation for large-scale structure surveys. 
Following the jackknife method, the total survey area is divided into 
$N$ subareas with similar effective area. In order to reduce the 
correlation between jackknife samples, the subareas need to be large enough
to include the modes at the scales we study.
For the `LowZ North (South)' survey area, we have $216$ ($90$) subareas in total
with each about $25\, \deg^2$. The partition is shown in \reffg{fig:jksample}. 
The survey area for `DR13 Group' is similar to `LowZ North' and divided into
$242$ subareas with $\sim 25\, \deg^2$ each.
By dropping one subarea at a time, we estimate the mean
pairwise momentum using the remaining $N-1$ subareas and obtain $N$ realizations.
The covariance matrix is then estimated from the jackknife samples as
\begin{align}
    C_{ij}^{\rm JK} = \frac{N-1}{N} \sum_{k=1}^{N}
    \left(p_i^k-\bar{p}_{i}\right)\left(p_j^k-\bar{p}_{j}\right),
\end{align}
in which $i$ and $j$ are the indices of the comoving separation bins and
$p^k$ is the pairwise momentum estimated using the $k$th jackknife sample.
We also apply the Hartlap factor~\cite{2007A&A...464..399H} to 
correct the biased inverse covariance matrix of jackknife samples
\begin{align}
    C^{-1} = \frac{N - N_{\rm bin} - 2}{N - 1}C^{-1}_{\rm JK},
\end{align}
The correlation coefficient matrix is estimated via the covariance matrix
\begin{align}
    R_{ij} = \frac{C_{ij}}{\sqrt{C_{ii}C_{jj}}}.
\end{align}

\begin{figure*}
    \centering
    \includegraphics[width=0.49\textwidth]{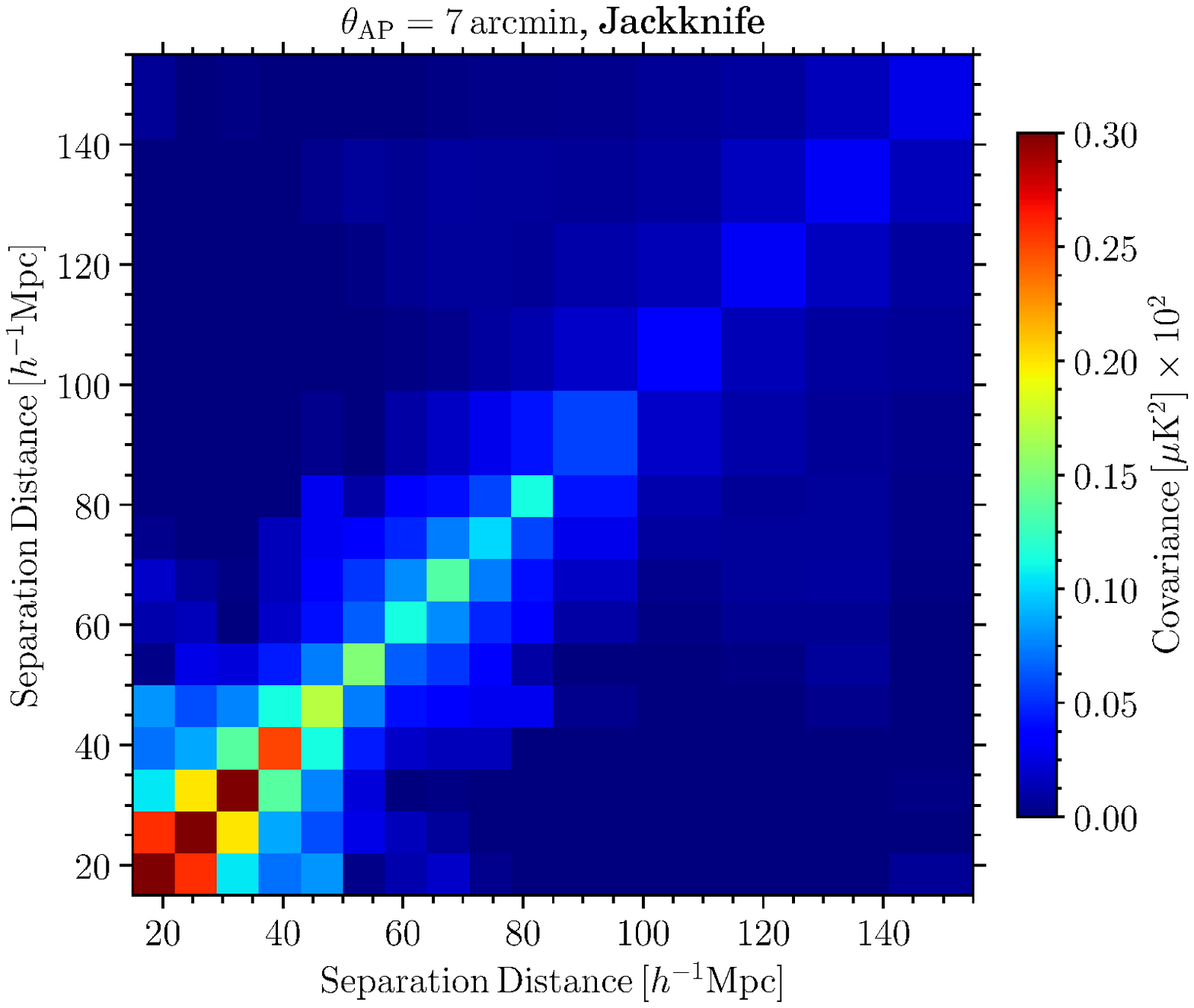}
    \hspace{-0.2cm}
    \includegraphics[width=0.49\textwidth]{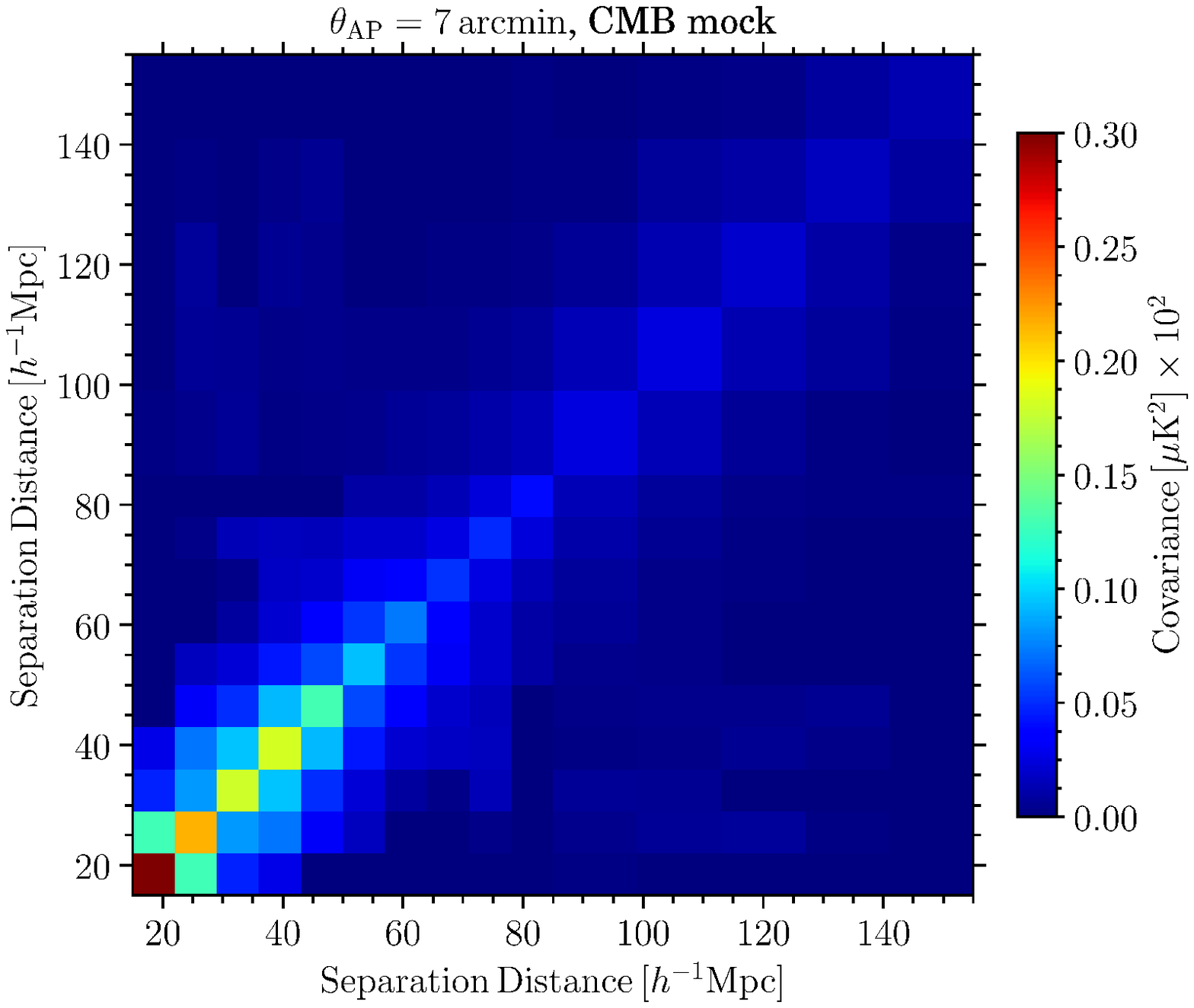}
    \\
    \hspace{-0.2cm}
    \includegraphics[width=0.49\textwidth]{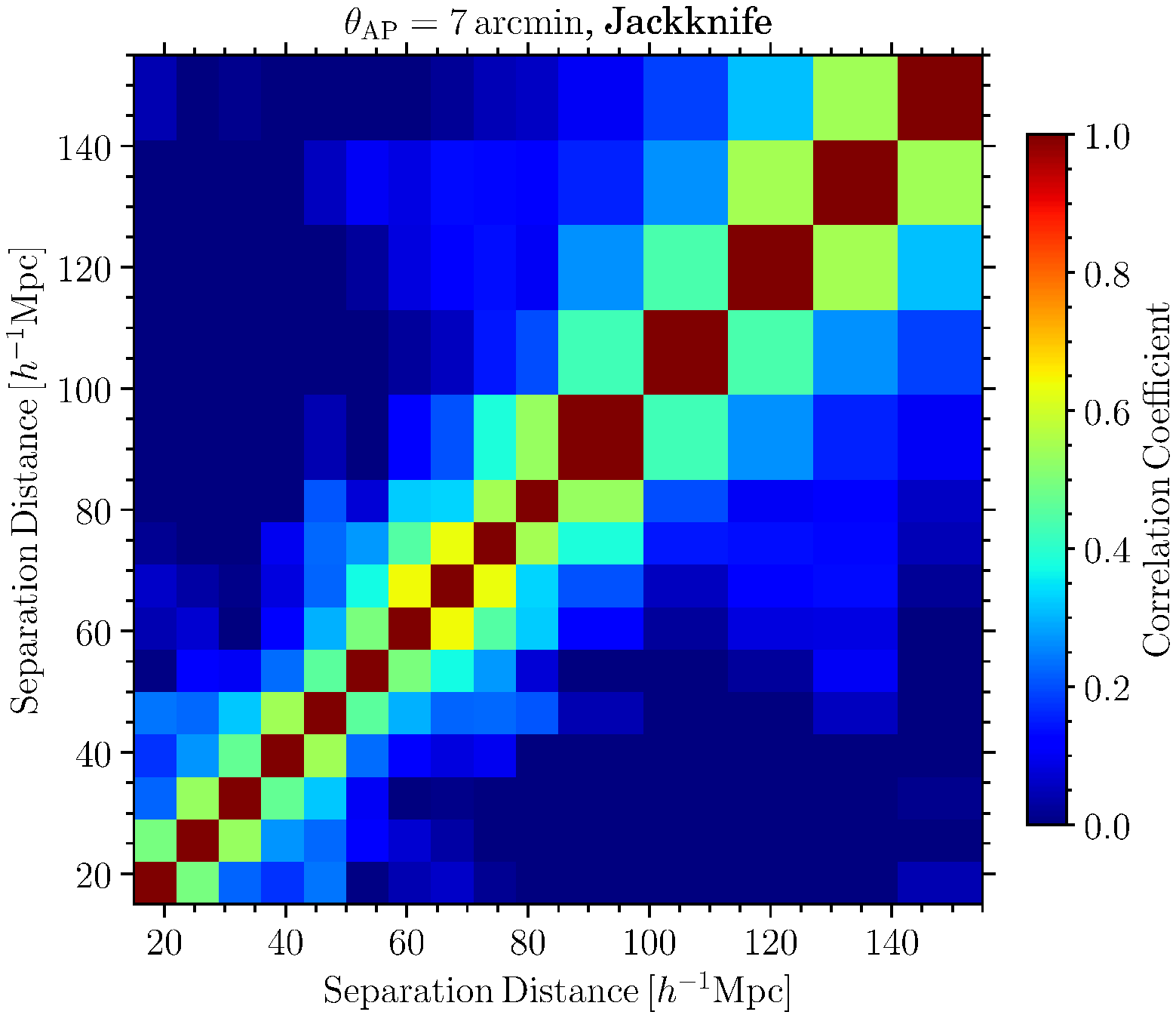}
    \hspace{-0.2cm}
    \includegraphics[width=0.49\textwidth]{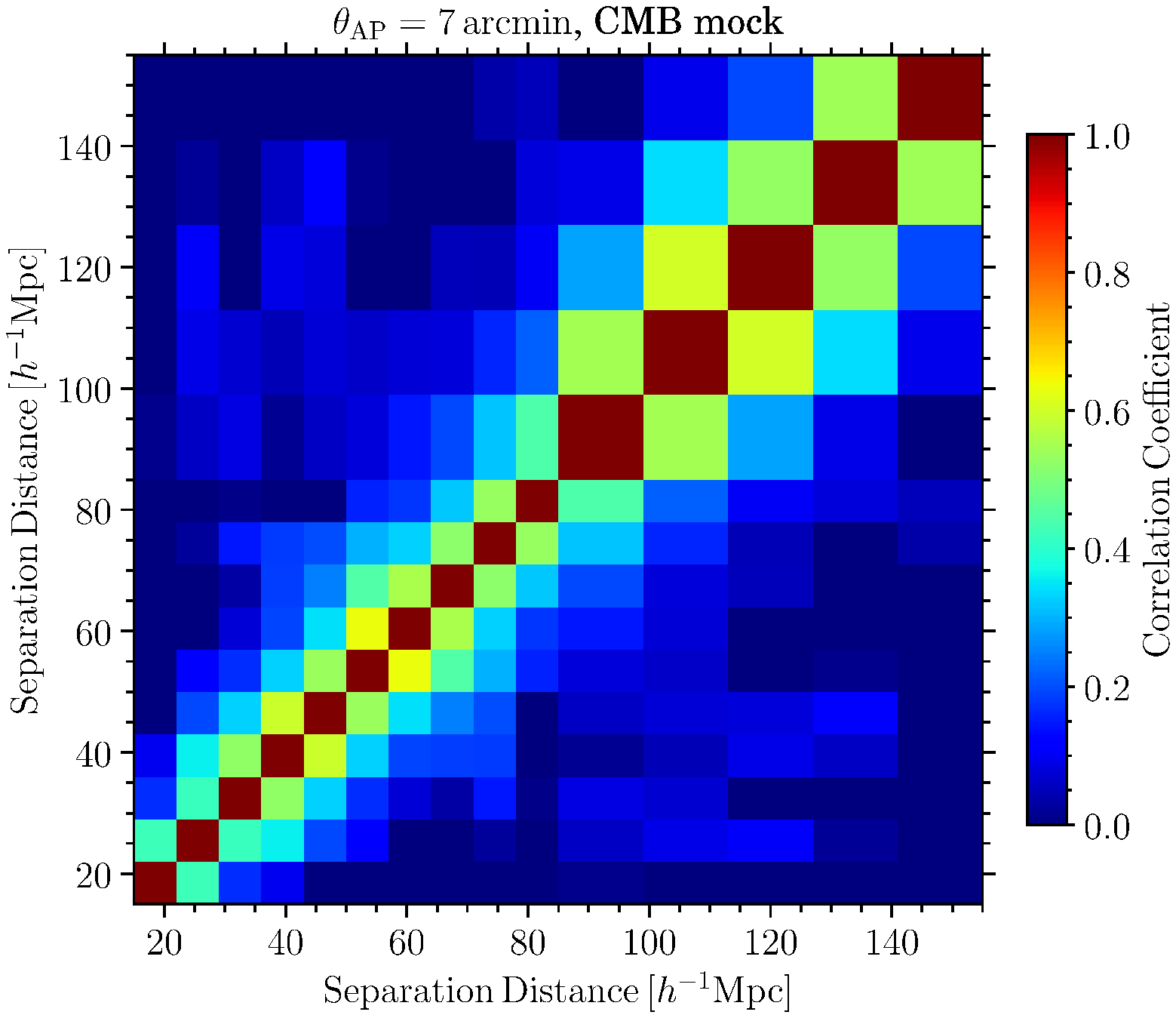}
    \caption{
        The covariance matrix ({\it the upper panels}) and correlation matrix 
        ({\it the lower panels}) for `LowZ North CGC'. The {\it left panels}
        show the covariance matrix estimated with the jackknife method, and the 
        {\it right panels} show the covariance matrix estimated with CMB mock samples. 
    }\label{fig:covcor}
\end{figure*}

The covariance matrices for the `LowZ North CGC' samples are shown in the
{\it upper panels} of \reffg{fig:covcor}, with the {\it left panel} showing the 
jackknife covariance and the {\it right panel} showing the CMB mock covariance. The `LowZ South CGC' and `DR13 Group' samples exhibit similar patterns in covariance matrix, but `LowZ South CGC' has slightly higher values due to the 
smaller survey volume, and `DR13 Group' has slightly lower values due to larger
samples. The {\it lower panel} of \reffg{fig:covcor} shows the correlation coefficient 
estimated with
jackknife samples ({\it left panel}) and CMB mock samples ({\it right panel}).
The higher correlations at larger comoving separations are recovered
with both jackknife and mock samples. The red and black step lines in~\reffg{fig:jk_vs_mock} show the diagonal error of the pairwise momentum in each bin, 
which is the square-root of the diagonal terms of the 
jackknife and CMB mock covariance matrices, estimated with `LowZ North CGC' samples. The errors are slightly underestimated
with the CMB mock samples, particularly at small separations, due to the lack of the
simulated kSZ signals, which would add measurement variance if included.
The green and blue step lines show the errors estimated with jackknife samples 
of `LowZ North CGC' and `DR13 Group', respectively.
We will use the covariance matrix and diagonal errors estimated from jackknife
samples for the rest of the analysis in this paper.

\subsection{Null tests}

\begin{figure*}
    \centering
    \small
    \begin{minipage}[t]{0.49\textwidth}
        \includegraphics[width=\textwidth]{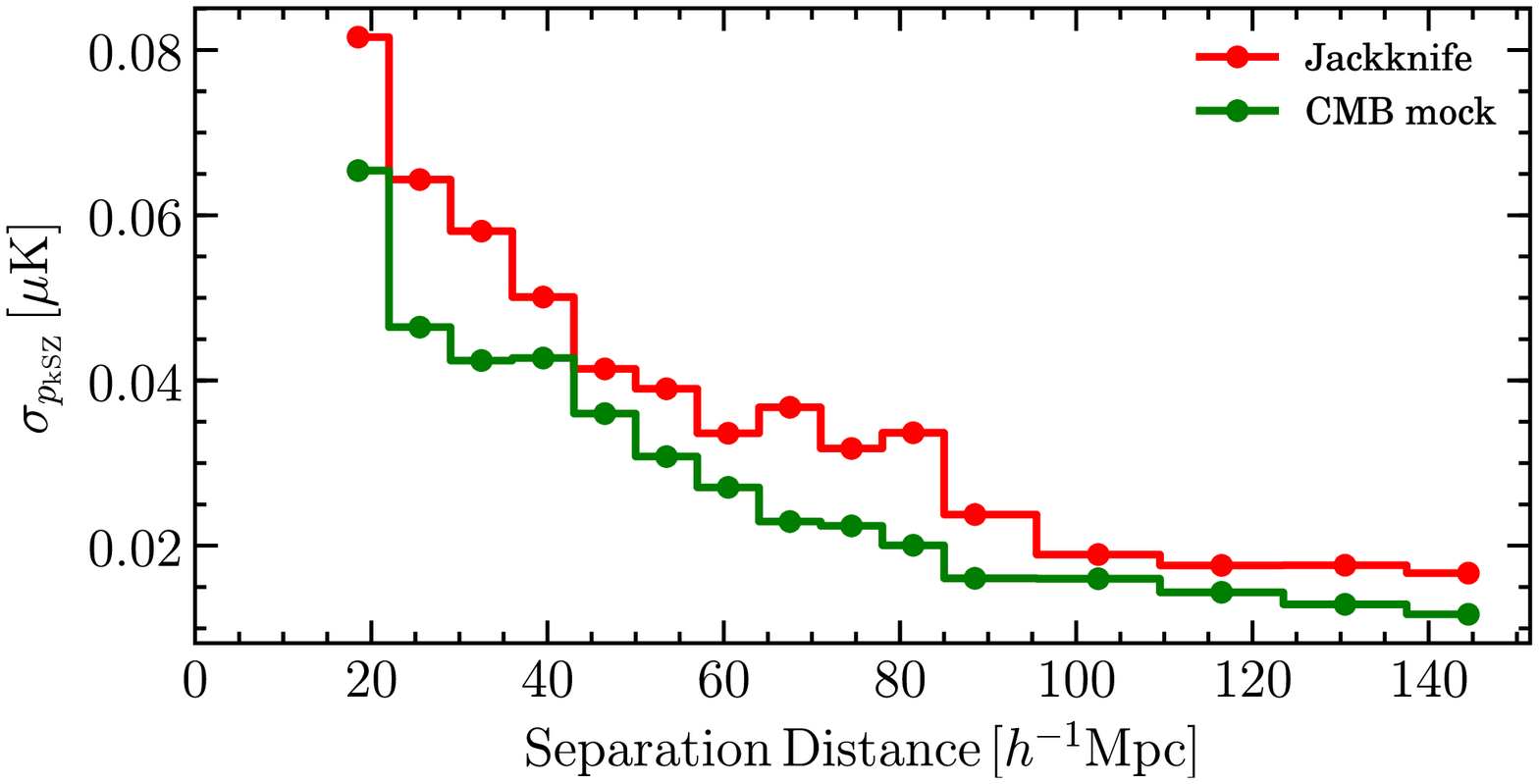}
        \caption{
            The errors estimated with the jackknife (red) and CMB mock (black) 
            samples for `LowZ North CGC'. The green and blue step lines show
            the errors estimated with jackknife samples 
            of `LowZ North CGC' and `DR13 Group', respectively.
        }\label{fig:jk_vs_mock}
    \end{minipage}\hfill
%
    \begin{minipage}[t]{0.49\textwidth}
        \includegraphics[width=\textwidth]{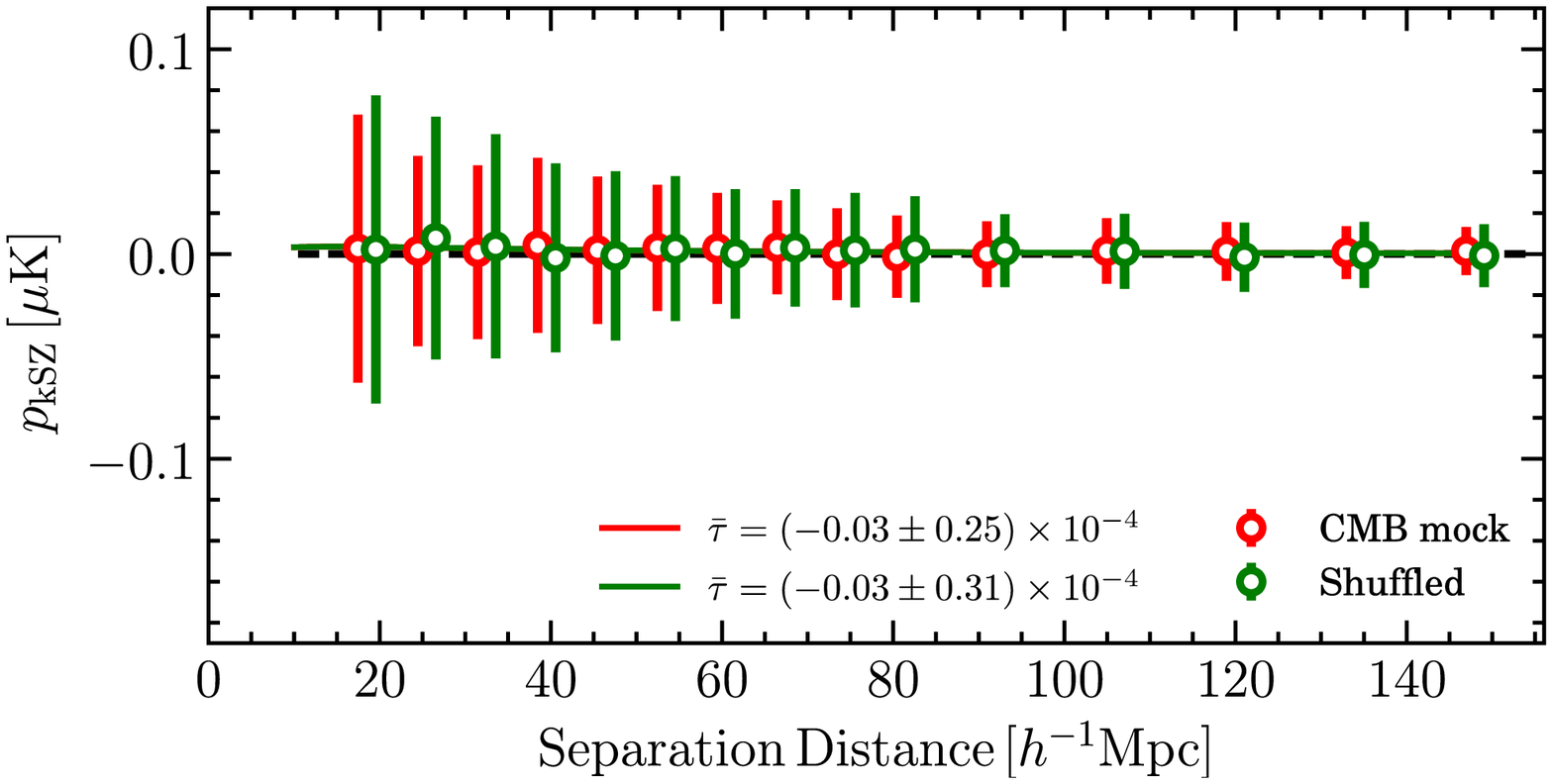}
        \caption{
            Results of the null test with shuffled (green) and mock (red) CMB maps.
        }\label{fig:null}
    \end{minipage}
\end{figure*}

In order to confirm that there is no systematic error in the maps or 
the analysis, we perform several null tests.
As mentioned above, the CMB mock maps that we used for the covariance 
matrix estimation, which contain no kSZ signal, provide a set of maps with which
we can perform a null test. By applying the 
estimator to the $200$ CMB mock maps, the fit to the averaged signal gives 
$\bar{\tau}_{\rm null} = (0.03\pm0.25) \times10^{-4}$, which is consistent 
with zero within $1\sigma$ C.L. The results are shown as the red points with error
bars in \reffg{fig:null}.

Another null test involves randomly shuffling the {\tt 2D-ILC}
CMB map, and applying the estimator to the shuffled maps. By averaging these
realizations, the signal is similar to the CMB mock samples, 
$\bar{\tau}_{\rm null} = (0.03\pm0.31)\times10^{-4}$, which is also consistent
with no signal. The results are shown as the green points with error bars in \reffg{fig:null}.

\subsection{The effect of AP filter size}

The optical depth is related to the detailed electron density profile
of the dark matter halo. Due to the AP filter, the measured optical depth,
$\bar{\tau}$, is the average value within the filter size. As shown 
in~\cite{2016A&A...586A.140P},
the measurement can achieve the maximum detection if the filter size matches 
the average angular size of the gas profile of dark matter halos.

\begin{figure}
    \centering
    \small
    \includegraphics[width=0.49\textwidth]{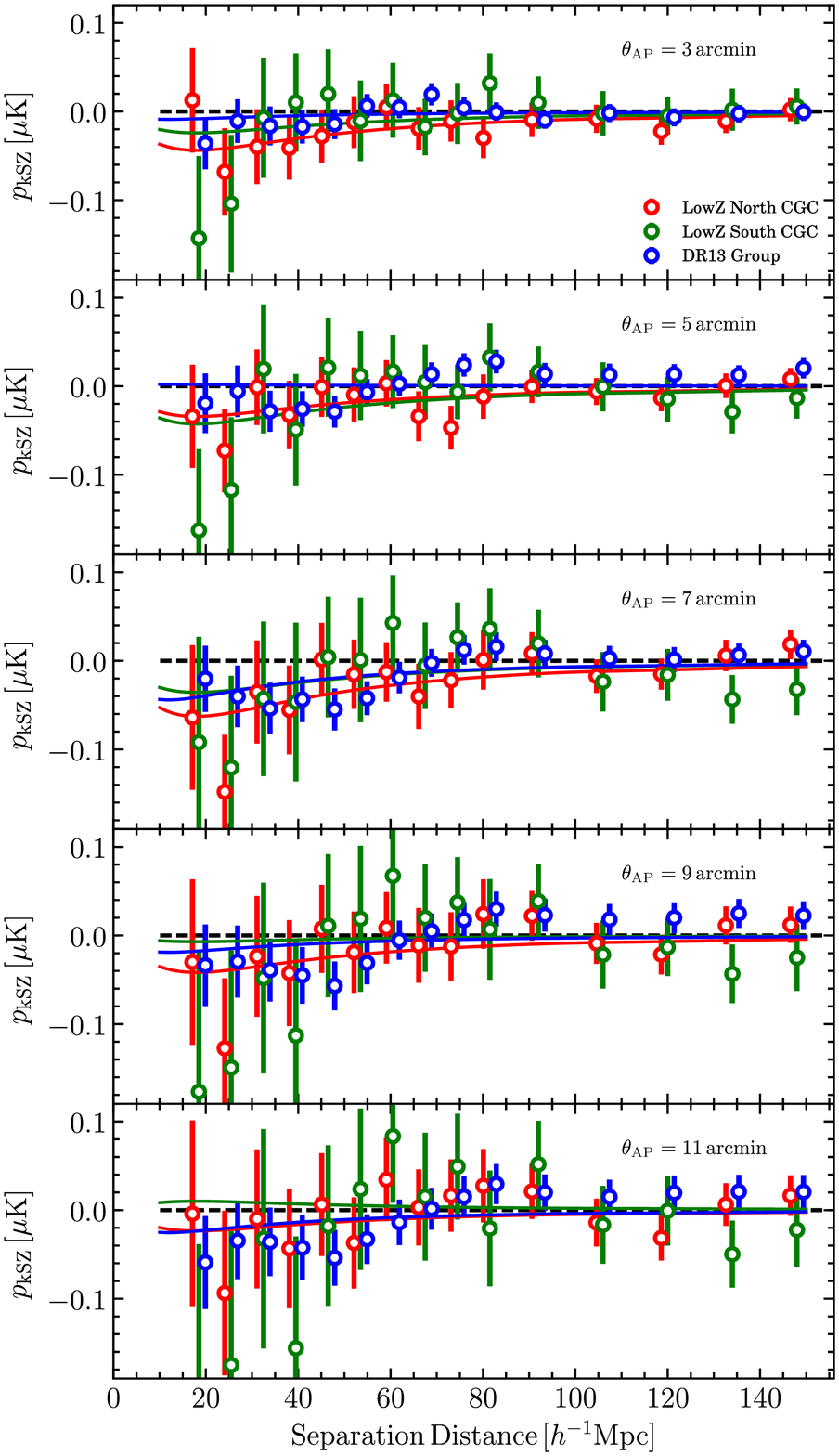}
    \caption{
        The measured $p_{\rm kSZ}$ with different AP sizes. From top to the bottom, 
        the AP radius is $3$, $5$, $7$, $9$ and $11\,{\rm arcmin}$.
        The error bars shown are the square-root of the diagonal terms of 
        the jackknife covariance matrix.
    }\label{fig:pkszap}
\end{figure}

\begin{figure}[h]
    \centering
    \small
    \includegraphics[width=0.49\textwidth]{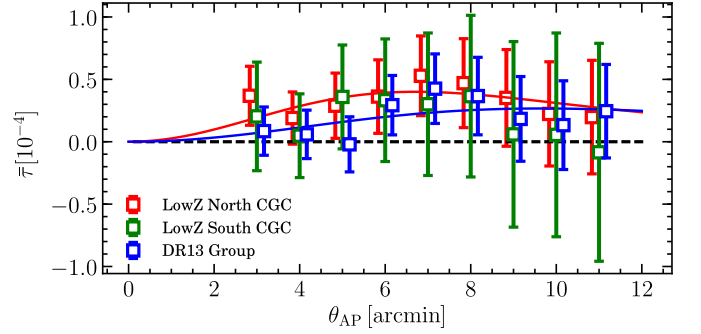}
    \caption{
        The best-fit $\bar{\tau}(\theta_{\rm AP})$ and its 
        $1\sigma$ errors with different AP size. 
        The solid lines show the best-fit theoretical prediction of 
        the mean optical depth as a function of $\theta_{\rm AP}$.
    }\label{fig:chisqap}
\end{figure}

We measured the $p_{\rm kSZ}$ functions with different AP filter sizes.
The AP radius, $\theta_{\rm AP},$ is varied from $3\,{\rm arcmin}$, 
which is approaching the resolution limit of the \planck CMB map, to $11\,{\rm arcmin}$.
The results with AP radii of $3$, $5$, $7$, $9$ and $11\,{\rm arcmin}$
are shown in \reffg{fig:pkszap}. The red, green and blue data points
show the results of the `LowZ North CGC', `LowZ South CGC' and `DR13 Group' 
catalogues respectively. 
The errors shown in the figure for the different bins are the square-root of the diagonal terms
of the jackknife covariance matrix.
Due to the smaller survey area and the sample size, 
the measurement with the `LowZ South CGC' catalogue has larger variance.
The theoretical predictions, shown as the solid line in \reffg{fig:pkszap}
are fitted to the measurements by minimizing
the $\chi^2$ equation in \refeq{eq:chisq}. 
The fitted values of $\bar{\tau}$ with different AP radii
are listed in \reftb{tab:tauap} and also shown as a function of $\theta_{\rm AP}$
in \reffg{fig:chisqap}.
A significant $\theta_{\rm AP}$ dependence can be found in these
the measurements.
The measured value of $\bar{\tau}$ peaks at an AP radius of $7\, {\rm arcmin},$ 
which is consistent with the detection in Fourier-space analysis
\cite{2017arXiv170507449S}, and decreases significantly when
the AP radius exceeds $8\, {\rm arcmin}$.

\begin{table*}
    \caption{
        The best-fit $\bar{\tau}$ with different AP radii. 
    }\label{tab:tauap}
    {\scriptsize
\begin{tabular}{C{1cm}|R{1.5cm}R{1cm}C{1.5cm}|R{1.5cm}R{1cm}C{1.5cm}|R{1.5cm}R{1cm}C{1.5cm}} \hline\hline
   $\theta_{\rm AP}$&\multicolumn{3}{c|}{LowZ North CGC}
   &\multicolumn{3}{c|}{LowZ South CGC}
   &\multicolumn{3}{c}{DR13 Group} \\

   $[{\rm arcmin}]$
   & $\bar{\tau}[10^{-5}]$ & S/N & $\chi^2/{\rm d.o.f.}$
   & $\bar{\tau}[10^{-5}]$ & S/N & $\chi^2/{\rm d.o.f.}$
   & $\bar{\tau}[10^{-5}]$ & S/N & $\chi^2/{\rm d.o.f.}$ \\\hline

   $3$
   & $3.7\pm2.4$ & $1.56\sigma$ & $0.52$
   & $2.0\pm4.4$ & $0.47\sigma$ & $0.31$
   & $0.9\pm1.9$ & $0.44\sigma$ & $0.77$\\

   $4$
   & $1.9\pm2.1$ & $0.91\sigma$ & $0.66$
   & $0.5\pm3.4$ & $0.14\sigma$ & $0.55$
   & $0.6\pm1.9$ & $0.30\sigma$ & $1.00$\\

   $5$
   & $2.9\pm2.6$ & $1.10\sigma$ & $0.80$
   & $3.6\pm4.2$ & $0.86\sigma$ & $0.55$
   & $-0.2\pm2.2$ & $-0.09\sigma$ & $0.84$\\

   $6$
   & $3.6\pm2.9$ & $1.23\sigma$ & $0.95$
   & $3.3\pm4.9$ & $0.68\sigma$ & $0.42$
   & $2.9\pm2.4$ & $1.23\sigma$ & $0.67$\\

   $7$
   & $5.3\pm3.2$ & $1.65\sigma$ & $0.81$
   & $3.0\pm5.7$ & $0.53\sigma$ & $0.46$
   & $4.3\pm2.8$ & $1.53\sigma$ & $0.53$\\

   $8$
   & $4.7\pm3.6$ & $1.32\sigma$ & $0.80$
   & $3.7\pm6.5$ & $0.56\sigma$ & $0.47$
   & $3.7\pm3.1$ & $1.18\sigma$ & $0.51$\\

   $9$
   & $3.5\pm3.9$ & $0.91\sigma$ & $0.57$
   & $0.6\pm7.4$ & $0.08\sigma$ & $0.49$
   & $1.8\pm3.4$ & $0.54\sigma$ & $0.63$\\

   $10$
   & $2.2\pm4.2$ & $0.53\sigma$ & $0.58$
   & $0.6\pm8.2$ & $0.07\sigma$ & $0.61$
   & $1.3\pm3.6$ & $0.37\sigma$ & $0.57$\\

   $11$
   & $2.0\pm4.6$ & $0.43\sigma$ & $0.58$
   & $-0.8\pm8.7$ & $-0.10\sigma$ & $0.75$
   & $2.4\pm3.7$ & $0.65\sigma$ & $0.42$\\

   \hline\hline
\end{tabular}

    }
\end{table*}

The mean optical depth as a function of AP radius can be modelled as
\cite{2017arXiv170507449S},
\begin{align}
    \bar{\tau}_{\rm m}(\theta_{\rm AP}) =
    \frac{\sigma_{\rm T} f_{\rm gas} M_{\rm h}}
    {\mu_{\rm e}m_{\rm p}D_{\rm A}^2(z_{\rm eff})} \int
    \frac{{\rm d}^2 \vec{\ell}}{(2\pi)^2}
    U(\ell\theta_{\rm AP}) N(\vec{\ell}) B(\vec{\ell}),
\end{align}
in which, $\sigma_{\rm T}=0.665\times10^{-24}\,{\rm cm}^2$ is the Thomson cross
section, $\mu_{\rm e}=1.14$ is the mean electron weight, $m_{\rm p}$ is the 
proton mass, $f_{\rm gas}=\Omega_{\rm b}/\Omega_{\rm m}\simeq0.155$ is the
cosmic mean fraction of gas, $M_{\rm h}$ is the halo mass and 
$D_{\rm A}(z_{\rm eff})$ is the angular distance at the effective redshift.

$U(\ell\theta_{\rm AP})$ expresses the AP filter function in Fourier space.
The real space AP filter function can be modelled as a step function,
\begin{align}
    U(\theta) = \frac{1}{\pi \theta^{2}_{\rm AP}} \times 
    \begin{cases}
        1 &\quad (0<\theta<\theta_{\rm AP}) \\
       -1 &\quad (\theta_{\rm AP}<\theta<\sqrt{2}\theta_{\rm AP}),
    \end{cases}
\end{align}
Then, in Fourier space, $U(\ell\theta_{\rm AP})$ is expressed as
\begin{align}
    U(x = \ell\theta_{\rm AP}) = 2 \left[W_{\rm top}(x) - W_{\rm top}(\sqrt{2}x) \right],
\end{align}
where $W_{\rm top}(x)=2J_1(x)/x$ is the top-hat smoothing window function expressed
using $J_1(x)$, which is the first order Bessel function of the first kind~\cite{Alonso16}. In the above $B(\vec{\ell})=e^{-\sigma_{\rm B}^2\ell^2/2}$ is the \planck beam function in
Fourier space, where $\sigma_{\rm B}={\rm FWHM}/\sqrt{8\ln2}$. The ${\rm FWHM}$ 
for the \planck {\tt 2D-ILC} CMB map we used is $5\,{\rm arcmin}$.

$N(\vec{\ell})$ describes the gas profile in Fourier space.
We assume that the gas profile can be expressed as a projected
Gaussian profile. In Fourier space, $N(\vec{\ell})$ is expressed as
\begin{align}
    N(\ell) = \exp\left[-\frac{\ell^2\sigma_{\rm R}^2}{2}\right],
\end{align}
where $\sigma_{\rm R}$ is the characteristic radius. 
A typical value of $\sigma_{\rm R}$ can be estimated from the halo radius $R$
divided by the angular diameter distance $D_{\rm A}$: 
$\sigma_{\rm R} = R/D_{\rm A}$. The halos can be identified as the spherical region 
where the mean density within the radius $R$ is $\Delta \rho_{\rm crit} E^2(z)$,
in which $\Delta$ is an experimental constant,
$\rho_{\rm crit}=2.775h^2\times10^{11}\,{\rm M}_{\rm \odot}{\rm Mpc}^{-3}$ 
is the critical density of the Universe today, 
and $E(z)=\sqrt{\Omega_{\rm m}(1 + z)^3 + \Omega_\Lambda}$.
If $\Delta=200$, the radius $R$ is close to the virial radius, and
$\sigma_{\rm R}$ is close to the virial angular size $\theta_{200}$,
\begin{align}\label{eq:sigmaR}
    \sigma_{\rm R} = \theta_{200} = \frac{1}{D_{\rm A}} 
    \left(\frac{3}{4\pi}\frac{M_{\rm h}}{\Delta\rho_{\rm crit}E^2(z)}\right)^{1/3}.
\end{align}

\begin{figure}
    \centering 
    \small
    \vspace{-0.3cm}
    \includegraphics[width=0.45\textwidth]{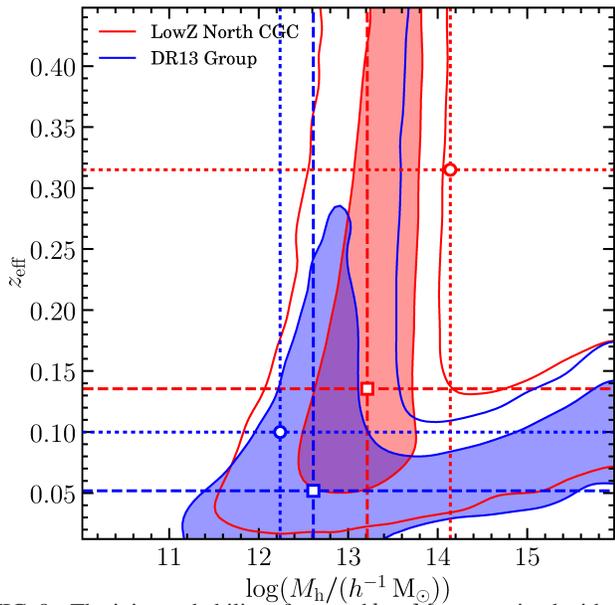}
    \vspace{-0.5cm}
    \caption{
        The joint probability of $z_{\rm eff}$ and $\log M_{\rm h}$ 
        constrained with the measured $\bar{\tau}$ of
        `LowZ North CGC' (red) and `DR13 Group' (blue).
        The filled inner and empty outer contours represent the $68\%$
        and $95\%$ C.L. respectively.
        The $68\%$ C.L. contour for `DR13 Group'
        goes beyond the parameter ranges of the plot.
        The squares with dashed cross lines indicate the best-fit
        values of $[z_{\rm eff}=0.15,\,\log M_{\rm h}=13.24]$ for
        `LowZ North CGC' and $[z_{\rm eff}=0.05,\,\log M_{\rm h}=12.58]$
        for `DR13 Group'. The circles with dotted cross lines
        indicate the median redshift and halo mass of the respective catalogues. 
    }\label{fig:contour}
    \vspace{-0.3cm}
\end{figure}


We assign $M_{\rm h}$ and $z_{\rm eff}$ as free parameters and fit
$\bar{\tau}_{\rm m}(\theta_{\rm AP})$ to the measurements by minimizing
the least-squares function,
\begin{align}
    \chi^2 = \sum_i (\bar{\tau}_{\rm m}(\theta_{{\rm AP},i})
    - \bar{\tau}(\theta_{{\rm AP},i}))^{2}/ \sigma_{\bar{\tau}}^2(\theta_{{\rm AP},i}).
\end{align}
We do not fit to the `LowZ South CGC' sample due to the lack of significance in measuring
$\bar{\tau}$ with this sample.
The joint probability of $z_{\rm eff}$ and $\log M_{\rm h}$
constrained with the measurements of `LowZ North CGC' and `DR13 Group'
are shown in \reffg{fig:contour}.
The filled inner and empty outer contours represent the $68\%$
and $95\%$ C.L. respectively.
The squares with dashed cross lines indicate the best-fit values.
For `LowZ North CGC' the best-fit values are
$[z_{\rm eff}=0.15,\,\log M_{\rm h}=13.24]$, while for `DR13 Group'
the best-fit values are $[z_{\rm eff}=0.05,\,\log M_{\rm h}=12.58]$.
The median redshift and halo mass of the catalogue are also
shown in \reffg{fig:contour} as the circles with dotted cross lines.
For `Group DR13' the median values are close to the best-fit values
and fall within the $68\%$ C.L. For the 
`LowZ North CGC', the median values fall slightly outside the $2\sigma$ contour,
which implies that the low-redshift and low-mass samples of
`LowZ North CGC' contribute more to the pairwise kSZ signal.
We will discuss the mass dependency further in \refsc{sec:mass_dependent}.
The theoretical prediction of $\bar{\tau}_{\rm m}(\theta_{\rm AP})$
using the best-fit $z_{\rm eff}$ and $\log M_{\rm h}$ values are shown as the solid lines
in \reffg{fig:chisqap}.

\subsection{The mean optical depth $\bar{\tau}$}

We quote the peak value of $\bar{\tau}$ at $7\, {\rm arcmin}$ as the
best-fit values, 
\begin{align}
    \bar{\tau}&=(0.53\pm0.32)\times10^{-4}\,(1.65\sigma) &
    {\rm LowZ\, North\, CGC}; \nmsk
    \bar{\tau}&=(0.30\pm0.57)\times10^{-4}\,(0.53\sigma) &
    {\rm LowZ\, South\, CGC}; \nmsk
    \bar{\tau}&=(0.43\pm0.28)\times10^{-4}\,(1.53\sigma) &
    {\rm DR13\, Group}. \nonumber
\end{align}
These results are consistent with the measurements in Fourier space
with the same catalogue~\cite{2017arXiv170507449S}, which used
the density-weighted pairwise kSZ estimator. By applying the analysis
in Fourier space, the method in~\cite{2017arXiv170507449S} avoids the correlations between different 
$k$-bins, thus achieving slightly higher significance.


\begin{figure}
    \centering
    \small
    \includegraphics[width=0.45\textwidth]{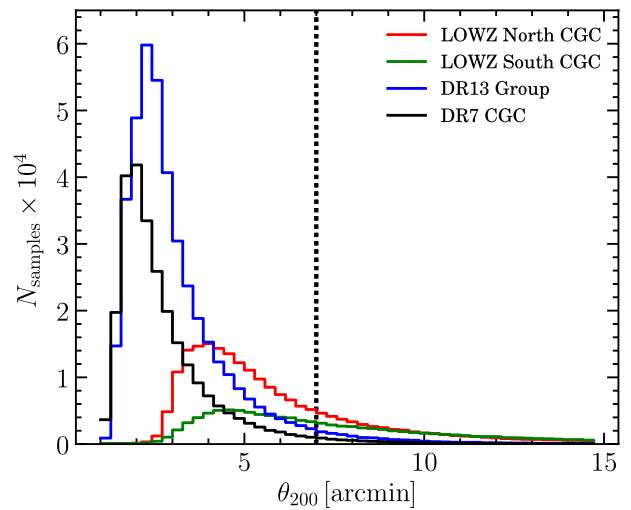}
    \caption{
        The $\theta_{200}$ distribution of the galaxy catalogues.
        The black dotted vertical line indicates $\theta_{200}=7\,{\rm arcmin}$.
    }\label{fig:theta200}
\end{figure}

The mean optical depth is also measured with different CMB maps.
\citepy{2017JCAP...03..008D} reports the results of 
$\bar{\tau}=(1.46\pm0.36)\times10^{-4}$, using the ACT CMB map and
BOSS DR11 galaxy catalogue. The angular resolution of the ACT CMB map
is much higher than \planck CMB maps. In their work, the $1.8\,{\rm arcmin}$
AP filter is applied to the ACT CMB map. Meanwhile, in order to 
avoid the systematic effect of the less massive clusters,
only the $20000$ most luminous sources are selected for the analysis.
Even more massive clusters are used in the analysis of \cite{2016MNRAS.461.3172S},
who report the measurement of $\bar{\tau}=(3.75\pm0.89)\times10^{-3}$
by using the SPT CMB map. Instead of the AP filter, the matched filter
is used in the analysis, with filter size of $0.5\,{\rm arcmin}$.
Because of the small radius of the filter, the mean optical depth reported in~\cite{2016MNRAS.461.3172S} is more sensitive to the central region of halos, where
the gas density is high compared to the outer regions. 
Figure~\ref{fig:theta200} shows the histogram of the virial angular
size, $\theta_{200},$ of each catalogue used in our study. The virial angular sizes 
are estimated via \refeq{eq:sigmaR}. The dotted vertical line shows the
filter radius of $7\,{\rm arcmin}$.
The filter size is large compared to the halo virial radius, which means that
the measured mean optical depth is averaged over large regions around the halo center.
As a result, our measured mean optical depth is lower.
However, our measurements of $\bar{\tau}$ are
consistent with previous analyses that used \planck CMB maps
\cite{2015PhRvL.115s1301H,2017arXiv170507449S}.

\subsection{Mass dependence}\label{sec:mass_dependent}

\begin{figure}
    \small
    \centering
    \includegraphics[width=0.49\textwidth]{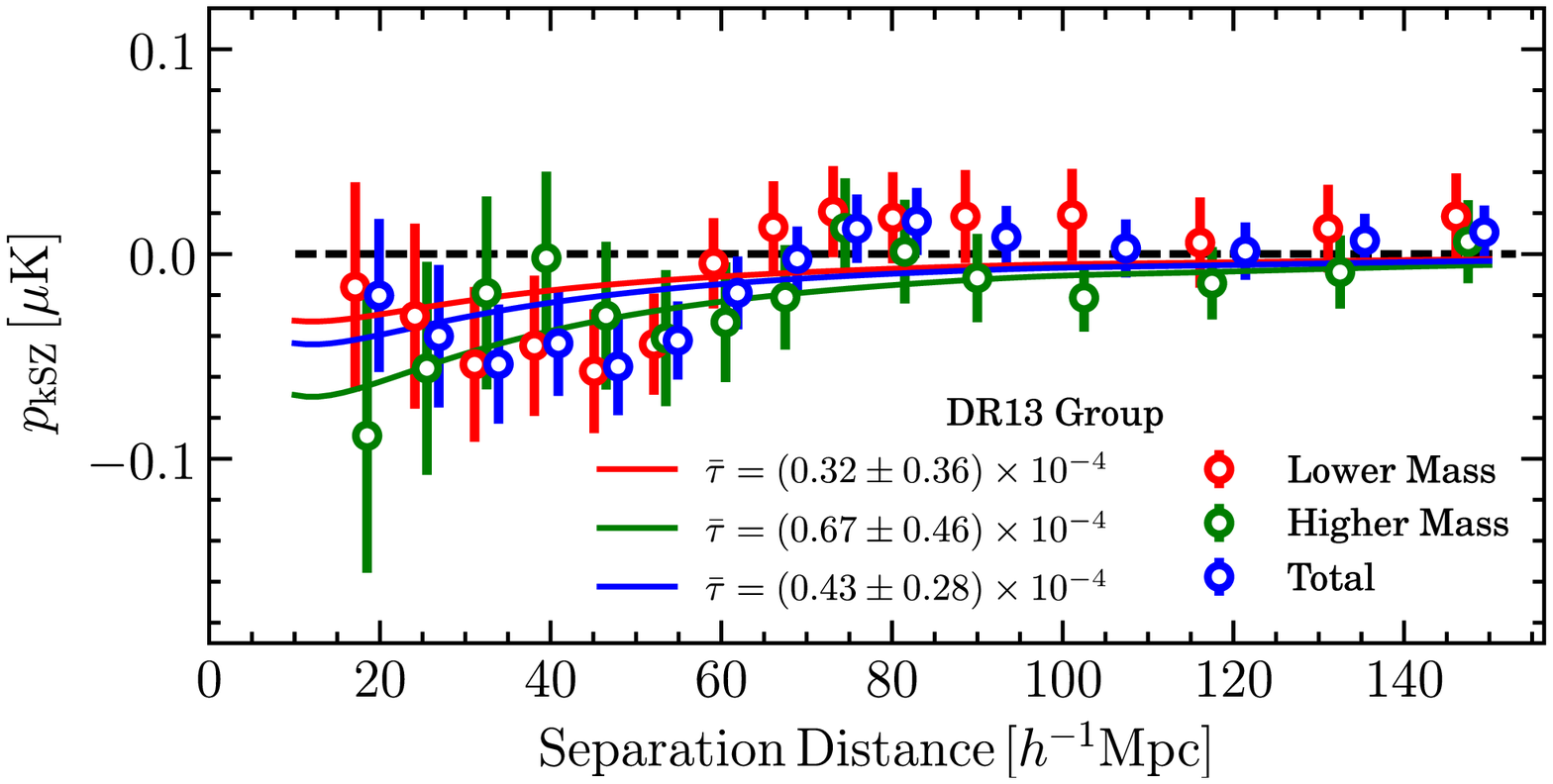}
    \includegraphics[width=0.49\textwidth]{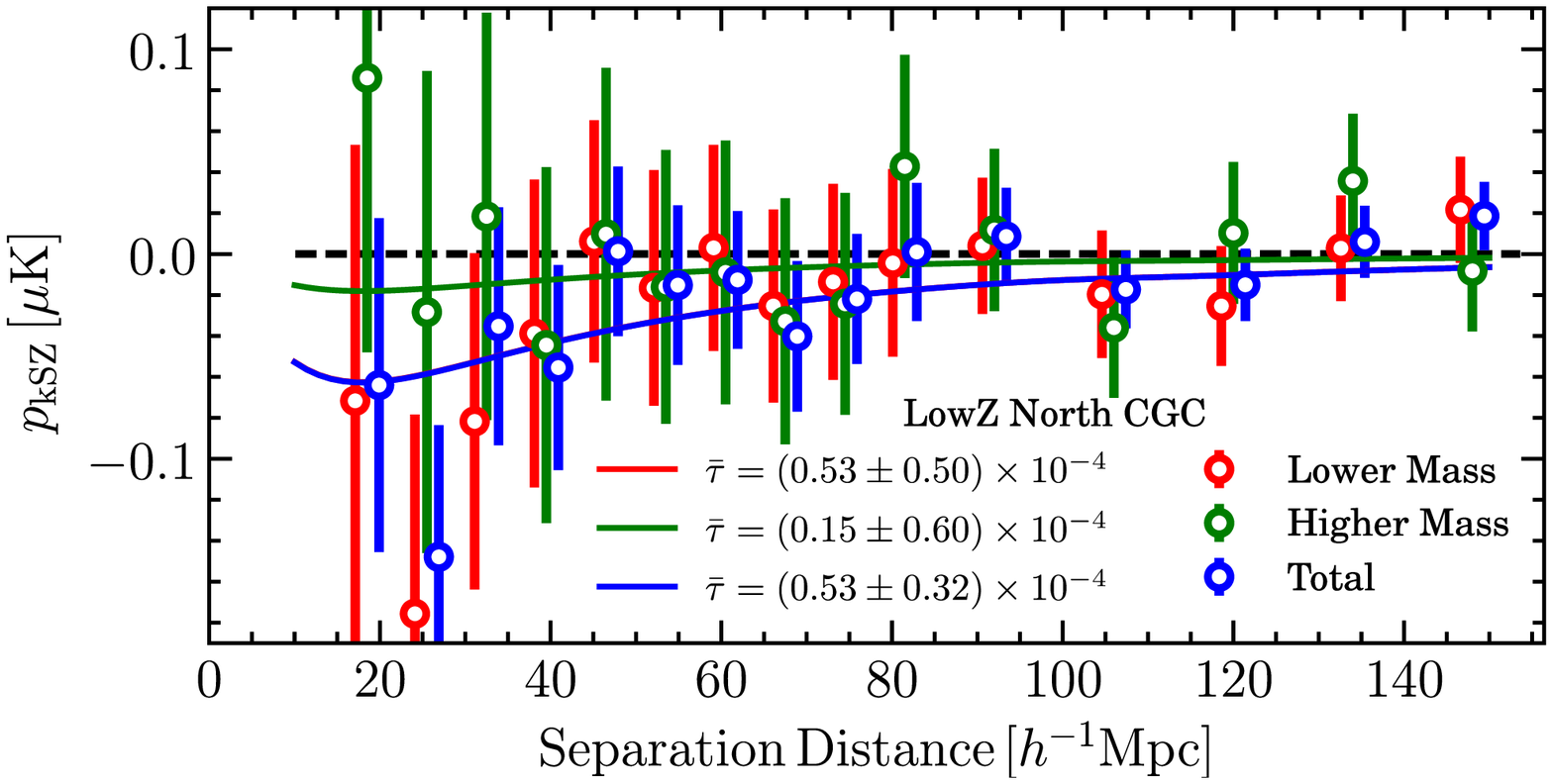}
    \caption{
        The measured kSZ pairwise momentum with different halo mass bins.
        The catalogues are split into higher and lower halo mass subcatalogues
        relative to the median value.
        The upper panel shows the results of the `DR13 Group' catalogue with
        median halo mass of $10^{12.24}\,h^{-1}{\rm M}_\odot$. The lower
        panel shows the results of `LowZ North CGC' with median halo mass
        of $10^{14.14}\,h^{-1}{\rm M}_\odot$. 
    }\label{fig:mass}
\end{figure}

\begin{figure}
    \small
    \centering
    \includegraphics[width=0.45\textwidth]{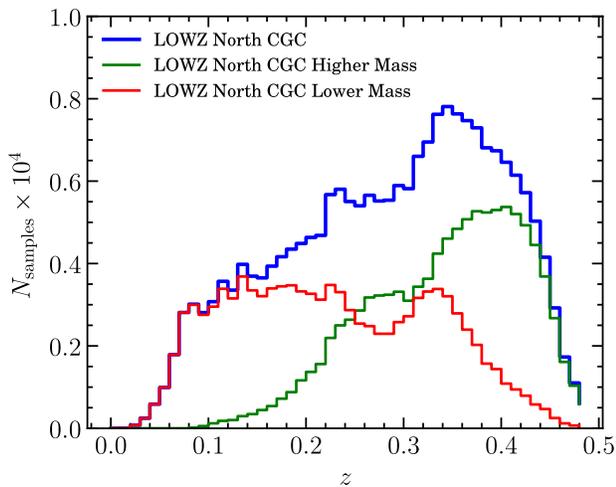}
    \caption{
        The redshift distributions of the higher/lower halo mass 
        subcatalogues of `LowZ North CGC'. 
    }\label{fig:masszhist}
\end{figure}

We split the `DR13 Group' catalogue into lower and higher halo mass 
catalogues according to the median value of the halo mass, 
$\log_{10}(M_{\rm h}^{\rm median}/(h^{-1}{\rm M}_\odot))=12.24$. 
By using the same pairwise estimator, we measured the mean optical depth
with the lower and higher halo mass catalogues. 

The top panel of \reffg{fig:mass} shows the results with the `DR13 Group'
lower mass (red), higher mass (green) and total (blue) catalogues. Since each subcatalogue
contains fewer samples, the errors are larger compared
to the full sample. However, for the subcatalogue of lower halo masses,
we measure the pairwise kSZ signal at $1\,\sigma$ C.L.
The best-fit mean optical depth of the lower mass halo catalogue is slightly smaller than
that of the higher mass halo catalogue, but still consistent within the $1\,\sigma$ error.

Similarly, we also split the `LowZ North CGC' catalogue into 
lower and higher halo mass catalogues.
The halo mass of the CGC samples are estimated according to the 
stellar-to-halo mass relation (SHMR) provided by \citepy{2012ApJ...744..159L},
and the stellar mass is obtained by matching the galaxy catalogue with
the Portsmouth SED-fit DR12 stellar mass catalogue\cite{2013MNRAS.435.2764M}
and the {\it GALEX-SDSS-WISE} Legacy Catalog (GSWLC)\cite{2016ApJS..227....2S}.
The galaxies without matched stellar mass information are ignored in the
mass-dependence analysis.
The median value of the `LowZ North CGC' halo mass is
$M_{\rm h}^{\rm median}=10^{14.14}\,h^{-1}{\rm M}_\odot$.
More than half of the sources in the `LowZ North CGC' catalogue are located in halos
with masses over $10^{14}\,h^{-1}{\rm M}_{\odot}$. The bottom panel of \reffg{fig:mass}
shows the results of each subcatalogue for `LowZ North CGC'.
We found that the higher mass subcatalogue lacks a measurement
of the pairwise kSZ signal, due to the high redshift range of
the high-mass halos.

Compared to the standard LRG criteria 
used in SDSS-I/II, the LowZ selection includes a bright magnitude cut, 
which excludes a significant number of low-redshift blue galaxies, 
but also excludes a fraction of bright galaxies in low-redshift
massive clusters~\cite{2015MNRAS.452..998H}. 
Figure~\ref{fig:masszhist} shows the redshift distribution of the
lower halo mass, higher halo mass and total `LowZ North CGC' subcatalogues.
It is clear that the distribution of high-mass halos is skewed towards higher redshifts.

In order to check whether the pairwise kSZ signal is dominated by 
the low-redshift samples, we split the `LowZ North CGC' catalogue
into two redshift subcatalogues relative to the median redshift, $z_{\rm median}=0.315$.
With the same pairwise kSZ estimator we find that, with the lower-redshift
subcatalogue, the mean optical depth is
$\bar{\tau} = (0.60\pm0.43)\times10^{-4}$, and with the higher-redshift
subcatalogue it is $\bar{\tau} = (0.28\pm0.58)\times10^{-4}$.
The results for $\bar{\tau}$ using this redshift split follow a similar trend to the measured $\bar{\tau}$ values for the `LowZ North CGC' lower and higher halo mass splits.

\section{Conclusions}


We presented the measurements of the pairwise kSZ momentum using the
\planck {\tt 2D-ILC} CMB map, which by construction has completely projected out the tSZ signal, 
and the Central Galaxy Catalogue (CGC) samples and group catalogue from BOSS DR12 and DR13.
The CGC is constructed by selecting the isolated galaxies from the
LowZ North/South catalogue of BOSS DR12. 
The group catalogue we used is based on the halo-based group finder
developed by \cite{2005MNRAS.356.1293Y} and recently updated with
the SDSS DR13 Northern Galactic Cap catalogue.

We used the pair-weighted pairwise kSZ estimator and
the AP filter to calculate the signal. The analysis was mainly focused on constraining the mean optical depth
$\bar{\tau}$. We first explored the AP filter radius size dependence of the
measurements and find that the radius of $7\,{\rm arcmin}$ gives the
maximum detection for $\bar{\tau}$. The results for the three catalogues are
\begin{align}
    \bar{\tau}&=(0.53\pm0.32)\times10^{-4}\,(1.65\sigma) &
    {\rm LowZ\, North\, CGC}; \nmsk
    \bar{\tau}&=(0.30\pm0.57)\times10^{-4}\,(0.53\sigma) &
    {\rm LowZ\, South\, CGC}; \nmsk
    \bar{\tau}&=(0.43\pm0.28)\times10^{-4}\,(1.53\sigma) &
    {\rm DR13\, Group}. \nonumber
\end{align}
We showed that the measured values of $\bar{\tau}$ are roughly consistent with the model values of $\bar{\tau}$ obtained using the median values of the halo mass and redshift of each catalogue.

Finally, we investigated the halo mass dependence by splitting the group 
catalogue and `LowZ North CGC' into higher and lower halo mass subcatalogues, 
according to their median halo mass. 
The group catalogue has most of the sources located in low-mass halos,
and the maximum halo mass for the lower-mass group subcatalogue is approximately
$10^{12}\, h^{-1}{\rm M}_{\odot}$. We achieved a $1\sigma$ C.L. detection with such a low-mass
catalogue.
In comparison, the galaxies in `LowZ North CGC' are mainly located in
high-mass halos, with a median value of $10^{14}\, h^{-1}{\rm M}_{\odot}$.
We achieved a similar detection with the lower mass `LowZ North CGC' subcatalogue,
compared to the full catalogue, but no detection with the higher mass `LowZ North CGC'  subcatalogue. This is because the galaxy samples located in high mass halos
are at higher redshifts, which make a smaller contribution to the total pairwise kSZ signals.

The study we performed here provides a viable method of probing the gas associated with central galaxies, which is an effective way to quantify the baryon fraction of these galaxies. We will perform such studies with future data when SDSS-IV data is released.

\acknowledgements
We thank Dr. Carlos Hern\'andez-Monteagudo for cross-checking the
results with DR7 CGC samples, Dr. Yu-Ting Wang for the discussion
on the BOSS LowZ/CMASS catalogues, Dr. You-Gang Wang for the
discussion on the DR13 Group Catalogue, Dr. Zhi-Gang Li for the
discussion on the error estimation, and Dr. Ming Li for the
discussion on kSZ measurements and CMB simulations. Y.C.L. and Y.Z.M. acknowledge the support from the National Research Foundation of South Africa with 
Grants No.~105925 and No.~104800. M.R. acknowledges funding from the European Research Council 
Starting Consolidator Grant No.~307209.

Computations were performed on the Hippo supercomputer at UKZN and the GPC supercomputer at the SciNet HPC Consortium.  
SciNet is funded by the Canada Foundation for Innovation under the 
auspices of Compute Canada, the Government of Ontario, 
Ontario Research Fund - Research Excellence, 
and the University of Toronto.

\bibliography{pksz_draft}

\end{document}